\documentclass[epsfig,floats,prl,twocolumn,preprintnumbers,floatfix,superscriptaddress]{revtex4-1}
\usepackage{graphicx}
\usepackage[latin2]{inputenc}
\usepackage{amsmath}
\usepackage{amssymb}
\usepackage{bm}
\usepackage{color}

\begin{document}
\title{Evolution of shear zones in granular packings under pressure}
\author{Mahnoush Madani}
\affiliation{Department of Physics, Institute for Advanced Studies in Basic Sciences, Zanjan 45137-66731, Iran}
\author{Maniya Maleki}
\affiliation{Department of Physics, Institute for Advanced Studies in Basic Sciences, Zanjan 45137-66731, Iran}
\affiliation{Optics Research Center, Institute for Advanced Studies in Basic Sciences, Zanjan 45137-66731, Iran}
\author{J\'anos T\"or\"ok}
\affiliation{MTA-BME Morphodynamics Research Group, Department of Theoretical Physics, Budapest University of 
Technology and Economics, Budapest H-1111, Hungary}
\author{M. Reza Shaebani}
\email{shaebani@lusi.uni-sb.de}
\affiliation{Department of Theoretical Physics $\&$ Center for Biophysics, Saarland University, 66041 
Saarbr\"ucken, Germany}

\begin{abstract}
Stress transmission in realistic granular media often occurs under external 
load and in the presence of boundary slip. We investigate shear localization in a 
split-bottom Couette cell with smooth walls subject to a confining pressure experimentally 
and by means of numerical simulations. We demonstrate how the characteristics of the 
shear zone, such as its center position and width, evolve as the confining pressure 
and wall slip modify the local effective friction coefficient of the material. For 
increasing applied pressure, the shear zone evolves toward the center of the cylinder 
and grows wider and the angular velocity reduces compared to the driving rate of the 
bottom disk. Moreover, the presence of slip promotes the transition from open shear 
zones at the top surface to closed shear zones inside the bulk. We also systematically 
vary the ratio of the effective friction near the bottom plate and in the bulk in 
simulations and observe the resulting impact on the surface flow profile. Besides 
the boundary conditions and external load, material properties such as grain size 
are also known to influence the effective friction coefficient. However, our numerical 
results reveal that the center position and width of the shear zone are insignificantly 
affected by the choice of the grain size as far as it remains small compared to the 
radius of the rotating bottom disk.
\end{abstract}

\maketitle
\section{Introduction}
\begin{figure*}
\begin{center}
\includegraphics[width=0.8\textwidth]{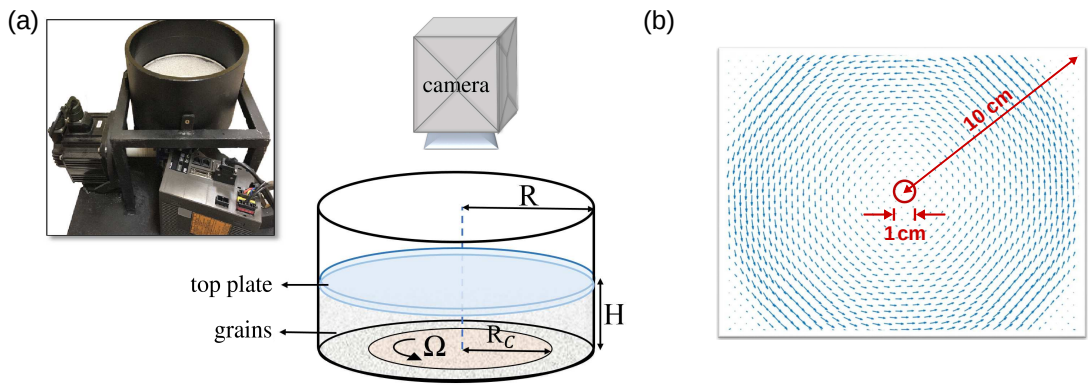}
\caption{(a) Sketch of the split-bottom Couette cell. (b) Displacement field of 
the grains from PIV analysis obtained from the subtraction of two successive 
configurations taken from the top surface of the granular pile. The red circle 
marks the region over which the axial angular velocity $\omega_{_0}$ is measured.}
\label{Fig:1}
\end{center}
\end{figure*}
Driven granular systems exhibit intriguing phenomena such as segregation, pattern 
formation, granular capillarity, and shear thickening and localization \cite{Kudrolli04,Aranson06,
Moosavi14,Fall10,Otsuki20,Schall10,Dijksman10,Fan17}. The latter is of great scientific, 
industrial, and geophysical importance. In slowly sheared granular materials, 
where lasting contacts play the main role in momentum exchange, one often observes 
a spatially inhomogeneous flow with the strain being localized in narrow shear 
zones \cite{Schall10,Dijksman10,Mueth00,Moosavi13,Fenistein03,Fenistein04,Cheng06}. 
The flow profile and shear stress are basically independent of the shear rate 
in this regime so that the shear localization cannot be fully captured by a 
constitutive stress-strain relation. In contrast to this regime, momentum 
transfer in fast granular flows is dominated by both lasting contacts and 
collisions. Here, the local flow rate can be characterized by a dimensionless 
inertial number which relates the shear stress to the shear rate, enabling to 
construct a stress-strain constitutive law and successfully describe the flow 
profile \cite{Jop06,Forterre08}.

When slowly shearing the boundaries of a granular system, narrow shear zones 
with a typical width of a few particle diameters form near the moving boundaries. 
The experiments on granular flows in Couette cells revealed that the shape of 
the wall-localized flow profile is independent of both filling height and shear 
rate and is solely determined by the characteristics of the granular microstructure 
\cite{Mueth00}. 

Nevertheless, shear zones away from the boundaries were also generated later 
in shear experiments in split-bottom Couette geometries, where the rough 
bottom plate of the cylindrical cell was split in two parts that rotate 
relative to each other \cite{Fenistein03,Fenistein04,Cheng06,Fenistein06}. In 
such geometries, the shear band is pinned in the bottom to the split position 
$r{=}R_c$ far from the side walls at $r{=}R$ (see Fig.\,\ref{Fig:1}). The shear 
zone is relatively wide as it takes advantage of the gravity to reach the top 
free surface through the bulk of the granular pile. In contrast to wall-localized 
shear zones, here the shape of the flow profile is governed by the filling height 
$H$. Additionally, the axial symmetry of the cylindrical split-bottom cell implies 
that the split position is another influential parameter in determining the 
position and width of the shear zone. For small driving rates $\Omega$ of the 
rotating bottom disk, the normalized flow profile is independent of $\Omega$. 
Starting from the split, the shear zone gradually grows wider and evolves 
toward the center of the cylinder with increasing height. Whether it reaches 
the top free surface depends on the ratio $H{/}R_c$; in deep granular layers 
the shear zone forms a dome-like shape in the bulk while in shallow layers it 
reaches the top free surface \cite{Cheng06,Fenistein06,Unger04,Torok07}. 

In shallow-layer experiments, it was shown that the axial angular velocity at 
the free surface $\omega_{_0}$--- defined as $\omega_{_0}{\equiv}\lim\limits_{r 
\to 0} \omega(r\,;h{=}H)$--- equals the driving rate $\Omega$ of the rough bottom 
disk \cite{Fenistein03,Fenistein04}. Moreover, the surface velocity profile $\omega(r)$ 
as a function of the radial coordinate $r$ is well fitted by an error function 
$\displaystyle\frac{\omega(r)}{\Omega}{=}\displaystyle\frac12\Big(1{-}\text{erf}
\left\{\frac{r{-}R_s}{\delta}\right\}\Big),$ with $R_s$ and $\delta$ being the 
center of the shear zone and its width, respectively. The center position follows 
$\displaystyle\frac{R_s}{R_c}{=}1{-}\Big(\frac{H}{R_c}\Big)^{5{/}2}$ and the 
width grows as a power-law $\delta{\sim}H^{2{/}3}$, i.e.\ faster than diffusion 
but slower than linear. A continuum constitutive model for dense granular flows 
was shown to capture the main features of the flow profile in shallow-layer 
experiments in the split-bottom Couette geometry \cite{Henann13,Li20}. However, 
axial slip occurs with increasing the filling height $H$ \cite{Cheng06,Fenistein06}; 
because of shearing between horizontal granular layers, $\omega_{_0}$ decreases 
and the velocity profile at the surface deviates from the error function shape. 
Above $H{/}R_c{\sim}0.6{-}0.7$, $\omega_{_0}{\rightarrow}0$ and the surface 
profile vanishes. 

So far, the Couette flow experiments to study shear localization in granular media 
have been carried out in cells with an open top boundary. Much less is known about 
the evolution of shear zones in more realistic media (as, for example, in geological 
processes) where the system is confined or subjected to an external pressure. Under 
such conditions, the dilation behavior and micro-structure of granular materials 
change \cite{Kobayakawa18,Shaebani12b,Bandi19,Das19,Shaebani12,Singh15,Murdoch13} 
which is expected to influence the flow profile. It was predicted based on an 
energy-dissipation variational approach that applying an external pressure should 
result in a similar behavior as observed with increasing filling height \cite{Unger04}.

Another point is that the role of effective friction coefficient near the bottom 
plate has yet to be resolved. In general, the effective friction in the bulk of 
the granular pile $\mu_{_\text{bulk}}^\text{eff}$ differs from the effective 
friction near the rotating disk $\mu_{_\text{bottom}}^\text{eff}$. One expects 
that the flow profile changes upon varying the ratio $\mu_{_\text{bottom}}^\text{eff}
{/}\mu_{_\text{bulk}}^\text{eff}$, since friction plays a key role in stress 
transmission in granular media \cite{Shaebani07,Goldenberg05,Shaebani09,GDRMiDi04,
Blair01} and particularly inside wide shear bands \cite{Luding07} (For a given 
$\mu_{_\text{bottom}}^\text{eff}{/}\mu_{_\text{bulk}}^\text{eff}$ ratio, the 
actual value of $\mu_{_\text{bulk}}^\text{eff}$ or $\mu_{_\text{bottom}}^\text{eff}$ 
does not influence the shear zone properties \cite{Unger04,Moosavi13}). It is 
however hard to tune the ratio in practice; even gluing a layer of grains to 
the walls with the same grains as in the bulk does not guarantee that the 
effective frictions are equal \cite{Moosavi13}. A systematic study of the 
parameters such as size and shape of the particles is necessary to better 
understand the role of material properties on the effective friction 
coefficient and shear deformation of granular materials. It has been shown 
that shearing of non-spherical grains in split-bottom geometries induces 
particle alignment (which reduces the effective friction) and leads to heap 
formation near the rotation axis \cite{Borzsonyi12,Wortel15,Fischer16}; 
however, the role of particle size is yet to be explored.

Furthermore, by considering rough boundaries in the previous experiments in Couette 
geometries, the possibility of wall slip has been often avoided. Nevertheless, wall 
slip occurs in real granular flows and it is not clear whether, and how, the 
presence of slip near the walls affects the flow profile. 

In this paper, we investigate the formation of shear band in a split-bottom 
Couette cell under external pressure $P$ and in the presence of slip between the 
grains and the bottom plate. By varying the weight of a top plate, we change the 
relative difference between the stress components of successive layers of 
grains, which affects the center position and width of the shear zone. The smooth 
bottom plate in our experimental setup allows for axial slip at the bottom layer, 
which significantly influences the strength of the surface flow profile and expedites 
the transition to closed shear zones inside the bulk. We also numerically study 
the role of the wall-to-bulk ratio of the effective friction coefficient as well 
as the particle size on shear localization. Therefore, by investigating the 
individual roles of the factors mentioned above, our main plan is to better 
understand the role of boundary constraints and conditions--- such as wall slip 
and roughness and applied loads--- on shear localization in granular flows.  

\section{Experimental Setup}
\label{Sec:Setup}
The experimental setup is similar to the split-bottom Couette cell previously 
used in experiments on granular flows \cite{Fenistein03,Fenistein04} (see 
Fig.\,\ref{Fig:1}). It consists of a rotating bottom disk of radius $R_c{=}
75\,\text{mm}$ and a stationary bottom ring that is attached to a cylinder 
of radius $R{=}100\,\text{mm}$. To apply an external pressure on the packing, 
we put a plexiglass plate with a mass ranging from $50$ to $550\,\text{gr}$ 
on top of the granular layer. For ease of comparison, we choose the pressure 
applied by a pile of grains with filling height $H$ as the reference pressure 
$P\!\!_{_0}$. Note that the applied pressure is in the order of the mean 
hydrostatic pressure of the granular pile and the anisotropic contact network 
(i.e.\ the network constructed by the normal vectors to contact planes) 
influenced by gravity is still far from being an ideal isotropic packing 
\cite{Shaebani09b}.

The cell is filled with grains up to height $H$. The average diameter of the 
glass beads is $2\,\text{mm}$ with size polydispersity of about $15\%$. This 
prevents crystallization effects near walls (i.e.\ shear-induced layering 
of equal-sized particles) that can affect the flow profile \cite{Mueth00,Chambon03}. 
The bottom plate and container walls are smooth, allowing for axial slip 
of the bottom layer of grains. The size of the gap at the split is less 
than $400\,\mu\text{m}$, which is much smaller than the minimal particle 
size, thus, no particle can escape.

\begin{figure}[t]
\centering
\includegraphics[width=0.48\textwidth]{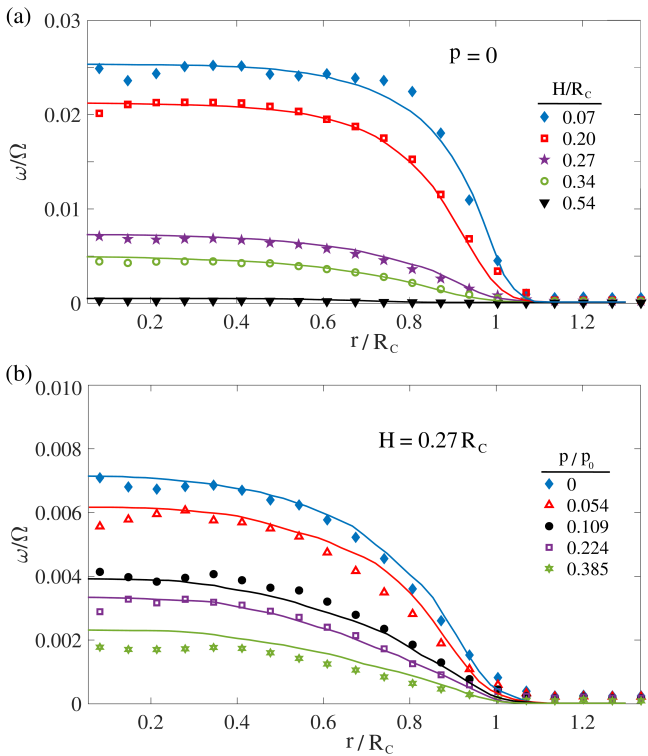}
\caption{Angular surface velocity $\omega$, normalized by the driving rate $\Omega$ 
of the bottom disk, versus the radial coordinate $r$, for (a) $P\!=\!0$ and various 
filling heights, and (b) $H\!=\!0.27\,R_c$ and different values of the external 
pressure. $P\!\!_{_0}$ is the reference pressure applied by the pile of grains with 
height $H$ for comparison. Symbols represent the experimental results and lines 
are obtained from the variational approach via Eq.\,(\ref{Eq:Variation}).}
\label{Fig:2}
\end{figure}

The bottom disk is rotated at angular velocity $\Omega$. The driving rate 
should be small enough to avoid rate-dependent stresses \cite{Hartley03}. 
The results reported here belong to $\Omega=0.13\,\text{rad}{/}\text{s}$.
After the flow reaches a steady state in a few seconds, we measure the flow 
profile at the top surface. The surface flow is monitored from above by a 
Casio camera with a CMOS sensor and frame rate $30\;\text{s}^{-1}$. The 
experiments are carried with and without the top plate. Because of using 
a camera with high pixel resolution of ${\sim}\,2.5\,\mu\text{m}$, the 
observation time of the surface flow that is required to reconstruct the 
particle displacement field is relatively short (ranging from a few seconds 
to a couple of minutes depending on the choice of $P$ and $H$ parameters). 
To obtain the mean angular velocity $\omega(r)$ at the top 
surface as a function of the radial coordinate $r$, particle image velocimetry 
method (PIV) is used, where the average angular cross-correlation function 
is determined in terms of $r$ for temporally separated frames. In order to 
obtain the axial angular velocity $\omega_{_0}$ at the top surface, we 
average the angular velocities within a small circle of radius $5\,\text{mm}$ 
around the cylinder axis; see the red marked circle in Fig.\,\ref{Fig:1}(b).

\begin{figure}[t]
\centering
\includegraphics[width=0.48\textwidth]{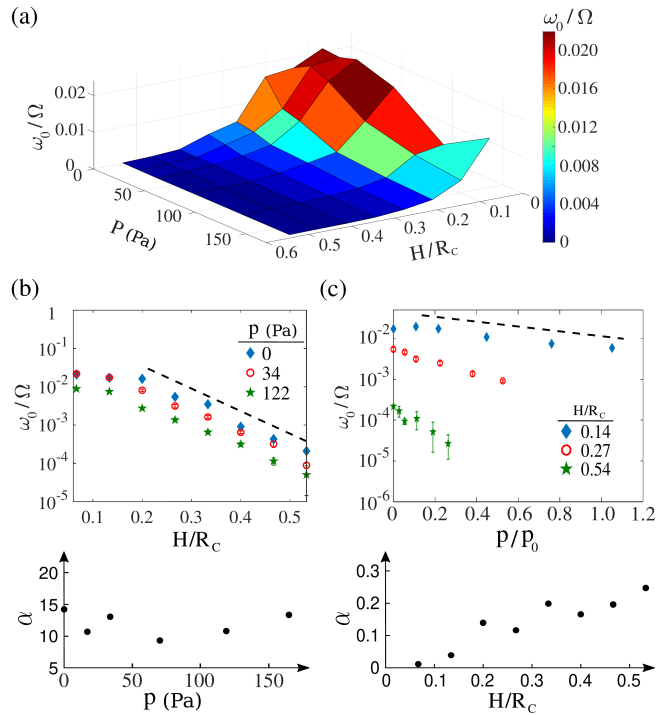}
\caption{(a) Axial angular velocity $\omega_{_0}$ in the ($H, P$) space. (b) 
top: $\omega_{_0}$ vs $H$ at different applied pressures. The dashed line 
shows an exponential decay as a guide to eye. bottom: Decay exponent 
$\alpha$ obtained from fitting an exponential function to the tail of 
$\omega_{_0}\!{-}H$ curves at different $P$ values. (c) top: $\omega_{_0}$ 
vs $P$ at different filling heights. bottom: Exponent $\alpha$ of the 
exponential decay of $\omega_{_0}\!{-}P$ tail at different filling heights 
$H$. The surface plot in panel (a) and symbols in top panels of (b) and 
(c) represent the experimental data.}
\label{Fig:3}
\end{figure}

\begin{figure*}
\centering
\includegraphics[width=0.99\textwidth]{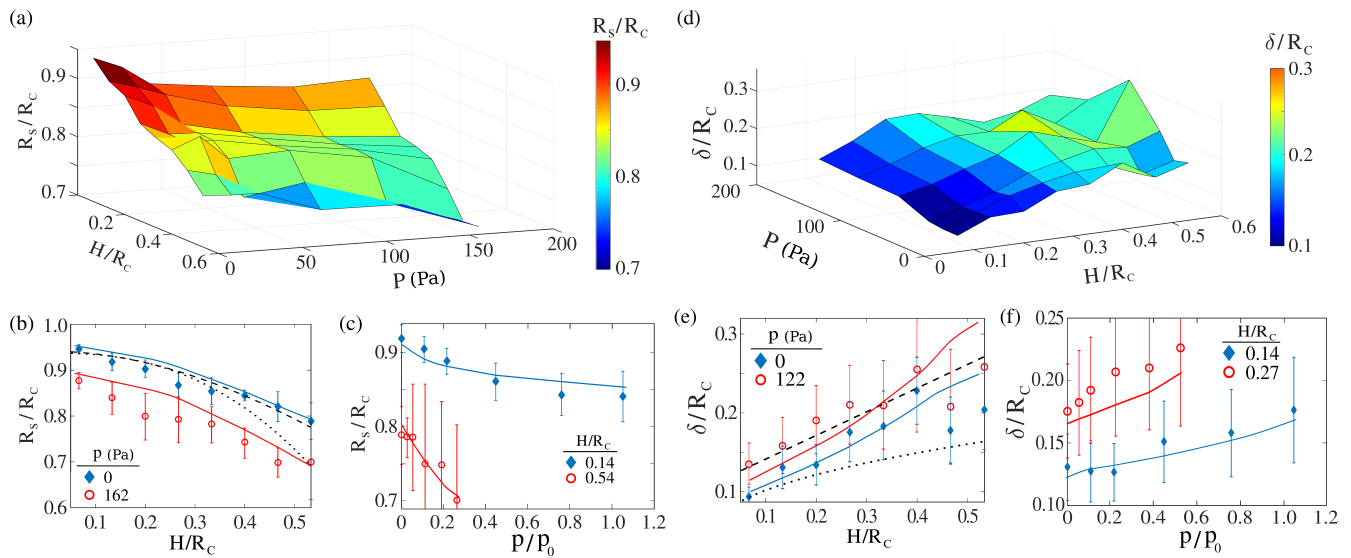}
\caption{(a) Shear zone position $R_s$ at the top surface in terms of the filling 
height $H$ and applied pressure $P$. (b) $R_s$ vs $H$ at different values of $P$. 
The dotted line represents $R_s{/}R_c{\propto}1{-}(\frac{H}{R_c})^{5{/}2}$ and the 
dashed line is a Gaussian fit to the free-surface (i.e.\ $P{=}0$) data. (c) $R_s$ 
vs $P$ at different values of $H$. (d) Shear zone width $\delta$ at the top surface 
in terms of $H$ and $P$. (e),(f) $\delta$ vs $H$ or $P$ for different choices of 
the other control parameter. The surface plots in panels (a) and (b) and symbols 
in all lower panels represent experimental results while solid lines are obtained 
numerically. The dashed and dotted lines in panel (e) correspond to $\delta{/}
R_c{\sim}H{/}R_c$ and $\delta{/}R_c{\sim}\sqrt{H{/}R_c}$, respectively.}
\label{Fig:4}
\end{figure*}

\section{Results}
\label{Sec:ExpResults}
First we investigate the role of wall slip on the velocity profiles. According 
to the previous experimental reports in split-bottom Couette cells with a rough 
bottom plate, no axial slip occurs in shallow layers ($H{/}R_c{\leq}0.45$) 
\cite{Cheng06,Fenistein04,Fenistein06}; the inner region above the central 
disk rotates as a solid while the outer part is stationary, and the angular 
velocity profile at the free surface is well fitted by the error function. 
At intermediate filling heights $0.45{<}H{/}R_c{\leq}0.65{-}0.7$, deviations 
from the ideal error function shape start to appear and the axial angular 
velocity $\omega_{_0}$ at the top surface decreases. The axial slip results 
from shear between horizontal layers inside the bulk, while the bottom layer 
still rotates at the same rate as the bottom plate because of the rough 
surface of the bottom disk \cite{Cheng06}. Eventually, the flow pattern 
at the free surface disappears for deep layers $H{/}R_c{>}0.7$, $\omega_{_0}$ 
approaches zero, and the shear zone is fully buried inside the bulk. We 
focus on the shallow-layer regime in our experiments with a smooth bottom 
plate. In the absence of the top plate, the flow profile at the free surface 
is shown in Fig.\,\ref{Fig:2}(a) for various filling heights. It can be seen 
that even at extremely shallow layers, the axial angular velocity $\omega_{_0}$ 
is nearly two orders of magnitude smaller than $\Omega$. This indicates that 
the majority of the slip between the stationary and the moving parts occurs 
at the bottom plate. It was shown in \cite{Moosavi13} that the smooth walls 
can be attractors of shear zones. $\omega_{_0}$ further decreases with 
increasing $H$ so that the surface flow practically disappears  for 
$H{/}R_c{\geq}0.45$. The shape of the surface flow profile also starts 
deviating from the error function at very shallow layers $H{/}R_c\!\simeq\!0.1$. 

The axial slip at the bottom smooth plate reduces the shear strength of the 
bulk material and weakens the flow profile. For the material properties used 
in our setup, we observe a considerable level of slip with $\omega_{_0}{/}
\Omega$ being around $0.03$ for extremely shallow layers. Still, the bulk 
is sheared strong enough so that the shear zone can proceed to the higher 
layers of grains. Then, an additional axial slip due to shear between horizontal 
layers inside the bulk further decreases $\omega_{_0}$ and the strength of 
the surface profile. As a result, the transition from shear zones reaching 
the free surface to closed shear zones inside the bulk occurs at smaller 
values of $H$ compared to the case of rough bottom walls. 

Next, we examine the surface flow profiles in the presence of the top glass 
plate. A few exemplary velocity profiles at $H{/}R_c\!=\!0.27$ and different 
external pressures $P$ are shown in Fig.\,\ref{Fig:2}(b). Increasing the applied 
pressure acts similarly to rising the filling height: it reduces $\omega_{_0}$ 
and weakens the strength of the surface flow. Fig.\,\ref{Fig:3}(a) shows the 
phase diagram of $\omega_{_0}$ in the ($H, P$) space. Approaching the plateau 
region at high $H$ or $P$ values, the shear zone is completely buried inside 
the bulk of the granular pile. A closer look at the behavior of $\omega_{_0}$ 
in Figs.\,\ref{Fig:3}(b),(c) reveals that the tail of $\omega_{_0}$ versus $H$ 
or $P$ decays nearly exponentially \cite{Cheng06} with a constant $\alpha$ that 
shows no systematic dependence on the pressure $P$ but gradually increases with 
$H$. Indeed, the relative difference between the components of the shear stress 
$\sigma_\text{ij}$ of two successive horizontal layers of grains depends on 
their height difference and is independent of the applied isotropic pressure 
or the absolute height of the pile above the two layers. That is why the decay 
of axial slip due to shear between horizontal layers is not affected by the 
choice of $H$ or $P$. Note that the increase of $\alpha$ with $H$ in 
Fig.\,\ref{Fig:3}(c) is caused by the increase of the reference pressure 
$P\!\!_{_0}$ with $H$.

Because of the strong slip at the bottom disk, shearing even a very shallow layer 
leads to $\omega_{_0}{<}\Omega$ and a modified flow profile at the top surface. 
Figure\,\ref{Fig:2} shows that with increasing $P$ or $H$ the profiles become 
more asymmetric with respect to the center position of the shear zone, but it 
is yet clear that the profile grows wider and the center position shifts toward 
the cylinder axis. An error-like function with four fit parameters was 
previously used to capture the asymmetric form of the flow profiles 
\cite{Fenistein06}. However, we checked that neither this form nor a logistic 
function provide a satisfactory fit to our data. Alternatively, we more 
precisely determine the characteristics of the shear zone from the shear rate 
information. For example, the center of shear zone can be characterized as the 
radial distance $R_s$ at which the shear rate has its maximum, i.e.\ $\text{d}
\epsilon_{r\theta}{/}\text{d}r|_{_{r{=}R_s}}{=}0$ with $\epsilon$ being the rate 
of the strain tensor. One can also quantify the width $\delta$ of the shear 
zone by, e.g., calculating the variance of the shear rate $\text{d}\omega{/}
\text{d}r$ around the peak at $r{=}R_s$. Both $R_s$ and $\delta$ parameters 
can be alternatively deduced from fitting the surface flow profile to an error 
function, which leads to qualitatively similar results and trends in terms of 
$P$ or $H$. Figures\,\ref{Fig:4}(a),(d) summarize the behavior of $R_s$ and 
$\delta$ in the ($H, P$) space. The relative decrease of $R_s$ versus $H$ is 
slower than the power-law $\frac{R_c{-}R_s}{R_c}{\propto}(\frac{H}{R_c})^{5{/}2}$ 
observed for shallow-layer shear in split-bottom Couette cells with a rough 
bottom disk \cite{Fenistein04}. In Fig.\,\ref{Fig:4}(b) we show a fit to the 
free surface ($P{=}0$) data as an example. Dramatic deviations can be observed 
for deeper layers. We find that a Gaussian decay $R_s{/}R_c\propto \text{exp}
\Big[-\displaystyle\frac{(H{/}R_c)^2}{2\,a^2}\Big]$ (with $a$ being a fit parameters) 
better captures the behavior [see the dashed line in Fig.\,\ref{Fig:4}(b)]. 
This holds also for other values of the external pressure as well as for $R_s$ 
vs $P$ curves at different filling heights. For the broadening of the shear zone with $H$, 
we find that $\delta$ grows faster than $\sqrt{H{/}R_c}$ but slower than $H{/}R_c$ for 
different values of the external pressure. The corresponding lines for the case of 
$P{=}0$ are shown in Fig.\,\ref{Fig:4}(e). Furthermore, when fitting $\delta{/}R_c$ 
vs $P$ to a power-law, the decay exponent again lies within $\left[0.5,1\right]$. 
It was similarly shown in rough-bottom geometries that the broadening of the shear 
zone at the free surface of the granular pile is faster than diffusion but 
remains slower than a linear growth \cite{Fenistein03}.

\section{Numerical approach}
In order to provide physical insight into how isotropic pressure influences the flow 
profile shape, we perform numerical simulations based on a variational minimization 
approach \cite{Onsager31,Onsager31b,Baker78,Einav06} that has been successfully 
applied to describe strain localization in granular materials \cite{Unger04,Moosavi13,
Torok07,Unger07,Szabo14,Borzsonyi09}. Despite the availability of efficient numerical 
tools for large-scale discrete element method simulations \cite{Shojaaee12,LAMMPS}, 
the variational approach is a computationally cheap way of obtaining the flow profile 
in symmetric geometries with simple boundary conditions. The main idea behind the 
variational approach is that the shear band forms along a path that minimizes the 
dissipation rate and complies with the boundary conditions. The cylindrical symmetry 
of the setup enables us to reduce the problem to a two-dimensional one. We consider 
a discrete 2D square lattice with a lattice size equal to the grain diameter. Denoting 
the radial coordinate of the shear band at height $h$ inside the bulk with $r(h)$, 
the boundary condition is $r(0)=R_c$. According to the original {\it narrow-shear-band} 
variant of the model \cite{Unger04}, the width of the shear band can be only one 
unit cell. Thus, two solid blocks slide past each other and the dissipation rate 
$\text{d}\gamma$ for a ring of sliding area with radius $r(h)$ is given by
\begin{equation}
\text{d}\gamma=\sigma_\text{tn}\,v(r)\,\text{d}s,
\label{Eq:DissipRate}
\end{equation}
where $\sigma_\text{tn}$ is the shear stress, $v(r){=}\Omega\,r(h)$ the sliding velocity 
between the blocks, and $\text{d}s{=}2\,\pi\,r(h)\,\text{d}l$ the sliding area 
($\text{d}l{=}\sqrt{(\text{d}h)^2{+}(\text{d}r)^2}$). To obtain the total dissipation 
rate, $\text{d}\gamma$ needs to be integrated over the whole surface of the shear 
band. Thus, up to a constant prefactor, the integral to be minimized can be 
expressed as
\begin{equation}
\displaystyle\int_\text{path}\!\!\!\!\!\!\!\!\!r^2\,\sigma_\text{tn}\;\text{d}l=\text{min},
\label{Eq:Variation}
\end{equation}
which results in an instantaneous narrow shear band. 
However, the material properties in the vicinity of the shear band is expected to 
change in practice. In a generalized {\it fluctuating-band} version of the model 
\cite{Torok07}, local fluctuations of the path due to random structural changes 
in the vicinity of the shear band have been introduced. As a result, the minimal path 
would slightly change in the next instance and the final shear profile is obtained 
by an ensemble average over all instantaneous narrow shear bands. Such a self-organized 
process allows for a finite width of the shear zone.  

\begin{figure}[t]
\centering
\includegraphics[width=0.47\textwidth]{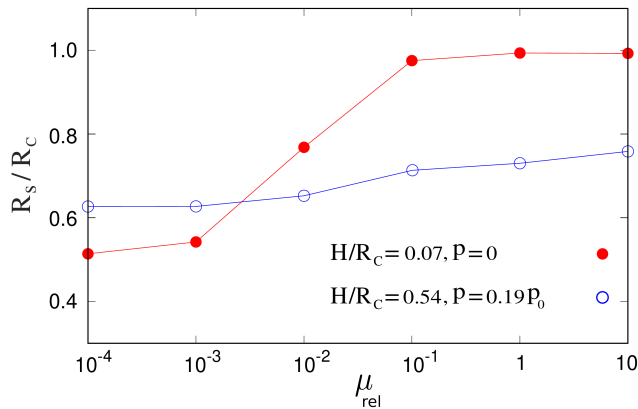}
\caption{Shear zone position $R_s$ at the top surface, scaled by the split position 
$R_c$, as a function of the relative friction coefficient $\mu_{_\text{eff}}$ for 
different values of $P$ and $H$. The results are obtained from the variational 
approach via Eq\,(\ref{Eq:Variation}).}
\label{Fig:5}
\end{figure}

We assume that the effective friction coefficient in the vicinity of the bottom plate 
$\mu_{_\text{bottom}}^\text{eff}$ differs from that of the bulk $\mu_{_\text{bulk}}^
\text{eff}$ in general. The ratio $\mu_{_\text{rel}}{=}\mu_{_\text{bottom}}^\text{eff}
{/}\mu_{_\text{bulk}}^\text{eff}$ is varied within $\left[0,10\right]$ in our simulations 
to investigate the role of axial slip at the bottom smooth plate. The initial and 
updated random strengths of the lattice sites follow the probability distribution 
$P(\mu^\text{eff})$ reported in Ref\cite{Shaebani08}. Because of the continuous 
agitation of the system, the Janssen effect plays no role here and we can assume a 
pressure $\rho\,g\,(H{-}h){+}P$ at height $h$ in the bulk. Then the shear 
stress is obtained as $\sigma_\text{tn}=\mu^\text{eff}\,\big(\rho\,g\,(H{-}h){+}P\big)$. 
Therefore, the isotropic pressure and the filling height influence the shear stress 
(and, thus, the variation integral) in a similar way. This explains why the same 
trends in terms of $P$ or $H$ are observed for the characteristics of the shear 
zone in our experiments (They are however not interchangeable because $H$ additionally 
sets the variational integration limit). To calculate the optimal path, most of the 
model parameters are fixed by the geometry and boundary conditions. The only remaining 
free parameters are the lattice size $d$ and the ratio of the effective friction 
coefficients of the bottom disk and bulk, $\mu_{_\text{rel}}$. We first choose $d{=}2
\,\text{mm}$ and tune $\mu_{_\text{rel}}$ as a free parameter to reproduce the experimental 
results and then study the role of grain size and wall slip by varying $d$ and 
$\mu_{_\text{rel}}$ parameters.  

Figure\,\ref{Fig:2}(a) shows that the flow profiles obtained by tuning the free parameter 
$\mu_{_\text{rel}}$ for each set of ($H, P$) parameters satisfactorily capture the trends 
observed in experiments. The resulting friction ratio $\mu_{_\text{rel}}$ slightly varies 
(within $\left[0.05,0.1\right]$), since variation of $H$ or $P$ could influence the 
arrangement and packing fraction of particles near the bottom plate and change the 
interparticle contact network (i.e.\ the positions of the contacts and their orientations). 
This leads to the modification of the effective friction coefficient $\mu_{_\text{bottom}}^\text{eff}$. 
The influence of $H$ or $P$ on the fitting parameter $\mu_{_\text{rel}}$ is more 
pronounced at shallow layers or low confining pressures. At higher values of $H$ 
or $P$, even fitting all flow profiles with a single optimal choice of $\mu_{_\text{rel}}$ 
leads to insignificant deviations from experimental trends. For example, all the flow 
profiles in Fig.\,\ref{Fig:2}(b) for a moderate filling height $H{/}R_c{=}0.27$ are 
obtained using $\mu_{_\text{rel}}{=}0.1$; the agreement is still satisfactory. 
We also checked that the error-function shape is recovered for $\mu_{_\text{rel}}{=}1$ 
in shallow layers, independent of the actual value of $\mu_{_\text{bulk}}^\text{eff}
{=}\mu_{_\text{bottom}}^\text{eff}$. The predicted center position and width of the 
shear zone shown in Fig.\,\ref{Fig:4} match well the experimental trends.

\begin{figure}[t]
\centering
\includegraphics[width=0.47\textwidth]{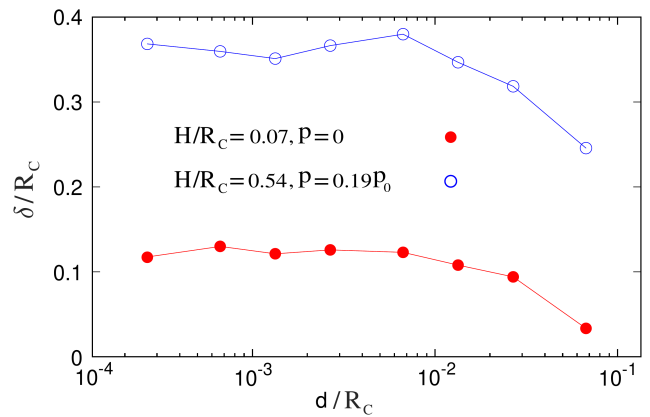}
\caption{Width of the shear zone at the top surface, scaled by the split position, 
in terms of the scaled grain size $d{/}R_c$ for different values of filling height 
and external pressure. The results are obtained from the variational approach via 
Eq\,(\ref{Eq:Variation}).}
\label{Fig:6}
\end{figure}

Next we systematically examine the role of effective friction at the bottom layer on 
shear localization, by varying $\mu_{_\text{rel}}{=}\mu_{_\text{bottom}}^\text{eff}
{/}\mu_{_\text{bulk}}^\text{eff}$ by several orders of magnitude. Overall, decreasing 
$\mu_{_\text{rel}}$ induces a stronger axial slip at the bottom disk and shifts the 
center position of the shear zone towards the cylinder axis. In the extreme regime 
of very shallow layers at low confining pressures, the inner region above a rough 
disk rotates as a solid and the center of the shear zone is almost above the split, 
i.e.\ $R_s{\simeq}R_c$. An evident transition in the center position towards the 
cylinder axis can be observed in this case upon decreasing $\mu_{_\text{rel}}$ (see 
Fig.\,\ref{Fig:5}). The change is much less pronounced for deep layers and high 
pressures, where the contribution of axial slip due to shear between horizontal 
layers of grains is significant. We observe no systematic dependence of the shear 
zone width on $\mu_{_\text{rel}}$.

Finally, we vary the lattice size $d$ in the variational calculations to see how 
grain size affects the position and width of the shear zone. By changing $d$ by more 
than two orders of magnitude while keeping the split position fixed at $R_c{=}75\,
\text{mm}$, we observe no dependence of the center position on the relative grain 
size $d{/}R_c$. Also the shear zone width is not affected in small grain size regime. 
However, the width decreases by increasing the coarse-graining size above $1\%$ of the 
container radius, as shown in Fig.\,\ref{Fig:6}. The energy dissipation cost of changing 
the narrow shear band path increases with the lattice size and eventually surpasses the 
gain by following the local changes in the material strength. This reduces the fluctuation 
range of the ensemble of instantaneous narrow shear bands for relatively large grains.

\section{Conclusion}
In conclusion we investigated shear localization in granular materials using a 
split-bottom Couette cell. While previous studies in similar geometries dealt 
with an open top boundary, here we confined the system by means of a top plate that 
applies an extra pressure on the granular pile, affecting the local stress and effective 
friction in the bulk of material. We studied the role of applied pressure on shear 
banding and demonstrated how shear zone characteristics, such as its center position 
and width, change upon varying the external load. We found that increasing external 
pressure promotes closed shear zones similarly to filling height. Moreover, previous 
studies mainly considered a simpler case of rough boundaries--- by gluing a layer of 
grains to the container walls--- to prevent slip. Here, we used smooth walls to allow 
for axial slip at the bottom plate and analyzed its influence on the emerged flow profiles. 
For a systematic study of the role of boundary roughness on granular rheology, we varied 
the ratio of the effective friction coefficients in the bulk and near the bottom plate 
in numerical simulations based on a variational approach to minimize the energy dissipation. 
We also numerically studied the role of grain size, as the material properties are known 
to influence the effective friction and response of granular materials in general 
and particularly the shear localization in Couette geometries.

Our results suggest that by tuning the boundary roughness and applied external pressure 
on a granular pile, the flow profile at a given filling height can be controlled to some 
extent. Understanding the rheology and mechanisms of shear localization in granular 
matter under confinement is crucial for industrial applications, e.g.\ in pharmaceutics 
and powder technology, as well as in geophysics. Shear localization can initiate 
avalanches, earthquakes, and faulting when the sheared layer withstands the weight 
of the upper layers. The results also once again highlight the applicability of the 
fluctuating-band variational approach with a self-organized random potential to 
describe granular shear flows. A possible application is the refraction of the shear 
zones when passing several layers with different confining pressures and material 
properties.  

\section{Acknowledgments}
We thank Matthias Schr\"oter and Wolfgang Losert for fruitful discussions and comments. 
J.T.\ acknowledges support by Hungarian National Research, Development and Innovation 
Office (NKFIH), under Grant No.\ OTKA K 116036, by the BME IE-VIZTKP2020.

\bibliography{Refs}

\end{document}